%% file: main.tex
\documentclass[conference]{IEEEtran}
\IEEEoverridecommandlockouts

\usepackage{cite}
\usepackage{amsmath,amssymb,amsfonts}
\usepackage{algorithmic}
\usepackage{graphicx}
\usepackage{textcomp}
\usepackage{xcolor}
\usepackage{balance}
\usepackage{microtype}
\usepackage{hyperref}
\def\BibTeX{{\rm B\kern-.05em{\sc i\kern-.025em b}\kern-.08em
    T\kern-.1667em\lower.7ex\hbox{E}\kern-.125emX}}
\begin{document}

\title{
ThreatGPT: An Agentic AI Framework for Enhancing Public Safety through Threat Modeling 
\vspace{-15pt}
}

\author{
\IEEEauthorblockN{\textsuperscript{} Sharif Noor Zisad}
\IEEEauthorblockA{\textit{Department of Computer Science} \\
\textit{University of Alabama at Birmingham}\\
Birmingham, Alabama, USA \\
szisad@uab.edu\vspace{-32pt}}
\and
\IEEEauthorblockN{\textsuperscript{}Ragib Hasan}
\IEEEauthorblockA{Department of Computer Science \\
University of Alabama at Birmingham\\
Birmingham, Alabama, USA \\
ragib@uab.edu\vspace{-32pt}}
}

\maketitle

\input{abstract}


\vspace{-5pt}
\input{introduction}
\vspace{-10pt}
\input{background}
\input{literature_review}
\vspace{-5pt}
\input{methodology}
\input{experiments}
\input{conclusion}
\balance
\bibliographystyle{IEEEtran}
\bibliography{references}

\end{document}

%% file: abstract.tex
\begin{abstract}
\label{sec:abstract}
As our cities and communities become smarter, the systems that keep us safe, such as traffic control centers, emergency response networks, and public transportation, also become more complex. With this complexity comes a greater risk of security threats that can affect not just machines but real people's lives. To address this challenge, we present ThreatGPT, an agentic Artificial Intelligence (AI) assistant built to help people whether they are engineers, safety officers, or policy makers to understand and analyze threats in public safety systems. Instead of requiring deep cybersecurity expertise, it allows users to simply describe the components of a system they are concerned about, such as login systems, data storage, or communication networks. Then, with the click of a button, users can choose how they want the system to be analyzed by using popular frameworks such as STRIDE, MITRE ATT\&CK, CVE reports, NIST, or CISA. ThreatGPT is unique because it does not just provide threat  information, but rather it acts like a knowledgeable partner. Using few-shot learning, the AI learns from examples and generates relevant smart threat models. It can highlight what might go wrong, how attackers could take advantage, and what can be done to prevent harm. Whether securing a city's infrastructure or a local health service, this tool adapts to users' needs. In simple terms, ThreatGPT brings together AI  and human judgment to make our public systems safer. It is designed not just to analyze threats, but to empower people to understand and act on them, faster, smarter, and with more confidence.
\end{abstract}

%% file: introduction.tex
\section{Introduction}
\label{sec:introduction}
Public safety systems form the backbone of modern urban infrastructure. They play an important role in maintaining societal stability by protecting populations during emergencies and supporting essential services such as transportation, healthcare, communication, and power distribution \cite{friedman2022public}. These systems are becoming smarter every day, thanks to digital technology and connected devices~\cite{chowdhury2025overview}. But as they become more advanced, they also become more vulnerable to cyber threats~\cite{wirtz2017cyberterrorism}. A small security weakness in one part of a public system can cause big problems. An attacker might break into a traffic control system and cause chaos on the roads. A failure in communication tools could delay the emergency response. In some cases, private information could be stolen or systems could be shut down. These risks are real and can affect both the systems and the people who depend on them. 

To protect public safety, we need to understand the threats before they happen. This process is called threat modeling~\cite{shostack2014threat}. It involves looking at the components of a system, figuring out what could go wrong, and finding ways to stop or reduce the impact of those problems. Many well-known models exist to help with this, such as STRIDE (Spoofing, Tampering, Repudiation, Information Disclosure, Denial of Service, and Elevation of Privilege)~\cite{khan2017stride}, MITRE ATT\&CK~\cite{xiong2022cyber}, CVE (Common Vulnerabilities and Exposures) databases~\cite{yosifova2021predicting}, NIST (National Institute of Standards and Technology) guidelines~\cite{ross2007managing}, and CISA (Cybersecurity and Infrastructure Security Agency) threat reports~\cite{czarnowski2022framework}. These models offer strong methods, but using them can be hard, especially for people who are not cybersecurity experts. Most threat modeling tools today are technical, rigid, or hard to customize. They do not easily fit different types of users or systems~\cite{theurich2023practices}. Many require deep security knowledge or manual work to apply the frameworks correctly. For a safety officer, developer, city planner, or healthcare manager, this can be overwhelming. There is a need for a tool that can make threat modeling easier, faster, and more human-centered.

To this end, we have developed ThreatGPT, an Agentic AI~\cite{acharya2025agentic, shavit2023practices} assistant designed to help people perform threat analysis in public safety systems. It uses the power of large language models to understand user input, analyze system components, and generate structured threat models based on popular frameworks. It not only gives general advice but also provides targeted, meaningful insights based on the system user describe. What sets ThreatGPT apart is its integration of Agentic AI capabilities with well-established cybersecurity frameworks to deliver a flexible, intelligent, and context-aware threat modeling experience. Unlike conventional tools that offer static checklists or require expert-level knowledge, ThreatGPT functions as an interactive assistant capable of understanding user-defined system components, guiding the analysis process, and adapting to varying levels of technical expertise. It leverages large language models and few-shot learning~\cite{parnami2022learning} to generate structured threat assessments while maintaining alignment with industry standard methodologies. This synthesis of generative AI~\cite{feuerriegel2024generative} with proven cybersecurity knowledge ensures that ThreatGPT delivers accurate, relevant, and actionable threat models tailored to diverse public safety systems. By enabling users to engage in an intuitive dialogue rather than navigate complex tooling, ThreatGPT democratizes access to high-quality threat analysis and empowers stakeholders from technical teams to public safety officials to proactively identify risks and strengthen system resilience.

Unlike traditional AI that only responds to commands, agentic AI is designed to act more like a helpful teammate. It can take initiative, ask questions, guide users, and adapt based on context. Instead of waiting for perfect instructions, an agentic AI works with the user thinking along the way. In the case of ThreatGPT, the AI does not just answer queries, it walks with the user through the full process of threat modeling, suggesting next steps, clarifying missing parts, and making sure nothing important is left out. This makes the experience more natural and supportive, especially for users without deep technical knowledge.

Every public safety system possesses a unique architecture, context, and risk profile. For instance, one user may describe the infrastructure of a hospital's IT network ~\cite{chowdhury2023security}, while another may focus on a smart traffic control system. Predefining and hardcoding threat modeling approaches for every potential use case is impractical and limits the flexibility of conventional systems. To address this variability, few-shot learning plays an important role. It enables the AI model to generalize from a small number of user-provided examples, allowing it to adapt dynamically to new system descriptions and generate relevant threat models without requiring extensive retraining or large labeled datasets.
\vspace{5pt} \\
\noindent\textbf{Contribution:}
The contributions of this paper are as follows:
\begin{enumerate}
    \item We trained an LLM with STRIDE, MITRE ATT\&CK, CVEs, NIST, CISA report to develop ThreatGPT, an Agentic AI system capable of generating extensive threat models for any system specified by the user.
    \item We demonstrated how ThreatGPT can support proactive threat analysis in public safety systems by helping users identify risks based on system components.
\end{enumerate}

\noindent\textbf{Organization:}
The structure of this paper is as follows: Section \ref{sec:introduction} covers the importance of threat modeling in ensuring public safety. Section \ref{sec:background} discusses the background of Agentic AI, Threat Modeling, and the framework we used to train our model, and Section \ref{sec:literature_review} reviews the related works. In Section \ref{sec:methodology}, the methodology of our proposed model is outlined along with the methodology. The experiments and the results are reviewed in Section \ref{sec:experiment}. Finally, Section \ref{sec:conclusion} presents the conclusion and discusses future directions.\\

%% file: background.tex
\section{Background}
\label{sec:background}
\vspace{-10pt}
\subsection{Agentic AI}
Agentic Artificial Intelligence~\cite{acharya2025agentic,shavit2023practices} is a new kind of AI that can take action on its own. It does not wait for step-by-step commands. Instead, it understands the user’s goal and helps complete tasks by reasoning and adapting. Unlike traditional AI, which only responds to direct input, agentic AI can guide users, ask questions, and suggest next steps. It acts more like a helpful assistant than just a tool. This makes it useful for solving problems that are complex or not clearly defined. The characteristics of Agentic AI~\cite{acharya2025agentic} are as follows:

\subsubsection{Goal driven}
These systems are designed with a clear objective in mind and consistently work toward achieving it. Unlike reactive models that only respond to specific prompts, a goal driven agent proactively pursues outcomes, making decisions and taking actions that align with its overall mission, even if the user does not explicitly request intermediate steps.

\subsubsection{Planning and Reasoning}
This model is capable of creating multi-step plans to achieve its goals. It does not just perform isolated tasks. It logically reasons through the problem, identifies the necessary steps, organizes them in sequence, and adapts its actions as it progresses, much like a human planning a project or solving a complex challenge.

\subsubsection{Autonomy}
One of the core features of Agentic AI is the ability to operate independently without needing constant human instructions. Once given an initial task or goal, an autonomous agent decides what to do next, manages sub-tasks, and resolves issues it encounters, much like a self-driving car making decisions on the road.

\subsubsection{Adaptability}
It can modify its actions and strategies based on new information or changing circumstances. If an obstacle arises or conditions change, the AI doesn’t freeze or fail. Instead, it intelligently adjusts its plan to continue pursuing its goal, ensuring resilience in dynamic environments.

\subsubsection{Memory and State}
It maintains memory of past interactions, decisions, and actions to inform its future behavior. This memory enables it to learn from previous experiences, avoid repetitive mistakes, build context over time, and deliver smarter, more consistent results instead of treating every request as completely new and isolated.

Agentic AI works by combining language understanding, memory, and learning. It can follow long conversations and remember what was said earlier. It can also handle missing or unclear information. Some agentic AI systems use few-shot learning~\cite{parnami2022learning}, where they learn from just a few examples. Others use reinforcement learning~\cite{wiering2012reinforcement}, where they get better by learning from feedback. These abilities help the AI improve over time and adjust to different users and situations.

This kind of AI is helpful in many real-world areas, such as healthcare, education, and public safety. In these fields, people often deal with systems that are large and complex. They need help making safe and smart choices. Agentic AI can support them by giving useful advice, checking for errors, and offering guidance. It helps users focus on what matters without needing deep technical knowledge. By doing so, agentic AI makes technology more human-friendly and easier to trust.

\subsection{Threat Modeling}
Threat modeling is the process of finding and understanding possible security problems in a system~\cite{shostack2014threat}. It helps people think about what could go wrong before it actually happens. By looking at the different parts of a system, we can find weak spots that an attacker might try to use. This process helps in building safer and more secure systems. It is an important part of designing systems that protect both data and people.

Threat modeling is useful in many fields, such as finance, healthcare, and public safety. It helps both experts and non-experts understand risks in simple ways. However, doing it manually can take a lot of time and effort. It often requires special knowledge and technical skills. That is why many researchers and developers are now exploring how artificial intelligence can support and improve the threat modeling process, making it faster, easier, and more accessible for everyone. The components of a threat model~\cite{zisad2024towards} are:

\subsubsection{Assets}
They are the critical components within a system that require protection from potential threats. These can include sensitive data, software applications, network infrastructure, intellectual property, and user credentials. Identifying assets is essential in threat modeling, as it helps prioritize security measures around the most valuable and vulnerable elements of the system.

\subsubsection{Entry Points}
These are the interfaces or access channels through which users, including potential attackers, can interact with a system. Examples of entry points include web portals, APIs, network ports, wireless connections, and physical access points. Properly identifying and securing entry points is crucial to prevent unauthorized access and reduce the system’s overall attack surface.

\subsubsection{Attacker Model}
It describes the characteristics, capabilities, and potential behaviors of adversaries who may target the system. This model considers factors such as the attacker’s motivation, resources, technical expertise, and level of access. A well-defined attacker model helps security analysts anticipate possible attack vectors and design more defensive strategies.

\subsubsection{Threats and Vulnerabilities}
Threats are potential events or actions that could harm a system, such as data breaches, denial of service attacks, or unauthorized privilege escalation. Vulnerabilities are the specific weaknesses or flaws, such as software bugs, misconfigurations, or insecure interfaces, that threats exploit. Understanding the relationship between threats and vulnerabilities is fundamental to assessing system risks and prioritizing mitigation efforts.

\subsubsection{Mitigation Strategies}
These involve implementing technical and procedural controls to reduce the risk from the identified threats and vulnerabilities. Examples include encryption, access control mechanisms, multi-factor authentication, intrusion detection systems, and regular software patching. Effective strategies help to minimize the likelihood of successful attacks and the potential impact on system operations.

\subsection{Threat Model Framework}
A threat model framework is a structured methodology used to systematically identify, evaluate, and address potential security threats and vulnerabilities within a system~\cite{bodeau2018cyber}. It provides a set of guidelines, processes, and classification schemes that help security analysts and system designers anticipate how an attacker might compromise assets, exploit entry points, and attack the system. The frameworks we used to develop our model are as follows.

\subsubsection{STRIDE}
It is a popular threat modeling framework created by Microsoft~\cite{khan2017stride}. It helps identify six types of security threats: Spoofing, Tampering, Repudiation, Information Disclosure, Denial of Service, and Elevation of Privilege. Each category focuses on a different kind of attack or weakness in a system. STRIDE is useful because it gives a simple way to think about how a system can be attacked. It works well during the design phase, helping analysts find threats early.

\subsubsection{MITRE ATT\&CK }
It is a large, open-source knowledge base of real-world cyber attacks~\cite{xiong2022cyber}. It shows the different steps attackers take, from getting access to stealing data. The framework is based on real tactics and techniques used by hackers. Each entry explains how an attack works and what tools may be used. MITRE ATT\&CK helps security teams understand attacker behavior and plan stronger defenses.

\subsubsection{CVE}
Common Vulnerabilities and Exposures (CVE) is a list of publicly known security issues~\cite{yosifova2021predicting}. Each CVE is a unique ID linked to a specific weakness in software or hardware. These IDs make it easy for people around the world to talk about the same problem. CVEs are widely used in tools, reports, and threat models to identify risks in systems. They are maintained by the MITRE Corporation~\cite{ambrose1990mitre} with support from the U.S. government.

\subsubsection{NIST}
The National Institute of Standards and Technology (NIST) provides guidelines and standards for cybersecurity~\cite{ross2007managing}. NIST helps organizations build strong and secure systems. Its publications, like the NIST Cybersecurity Framework~\cite{white2022nist}, are widely used in both the government and private sectors. NIST offers best practices on risk management, access control, encryption, and more. These guidelines are helpful during system design, development, and security evaluation.

\subsubsection{CISA}
The Cybersecurity and Infrastructure Security Agency (CISA) is a U.S. government agency that protects critical infrastructure~\cite{czarnowski2022framework}. It publishes regular reports on threats, vulnerabilities, and response strategies. CISA shares alerts, best practices, and risk assessments to help public and private organizations stay secure. Its reports are trusted sources for current cybersecurity trends and threats. CISA also works with other agencies to respond to major cyber incidents.

\subsubsection{NVD}
The National Vulnerability Database (NVD) is a detailed resource that builds on CVEs and is managed by NIST~\cite{booth2013national}. It adds more information to each CVE, such as severity scores, fix status, and impact details. NVD helps security professionals understand how serious a vulnerability is and how it affects systems. It is often used in automated security tools for scanning and threat detection.

%% file: literature_review.tex
\section{Literature Review}
\label{sec:literature_review}
In this section, we will discuss some recent works using AI for threat analysis. 

Hamza et al. highlighted the growing need for proactive cyber-risk management in a world where threats are constantly evolving~\cite{rauf2025using}. Traditional methods like rule-based and signature-based detection are no longer enough, as they struggle to handle new and advanced attacks. They explored how Generative AI models, such as GANs and VAEs, can be used to simulate realistic and unseen cyberattack scenarios, helping organizations test and improve their defense systems. This shift from reactive to proactive threat modeling allows for better preparation against complex threats such as advanced persistent threats (APTs). They also emphasized the value of combining conventional security methods with AI-driven techniques to build adaptive and resilient cybersecurity systems.

Mohamed et al. proposed a comprehensive framework for cyber threat detection by leveraging AI, natural language processing (NLP), and malware analysis ~\cite{mohamed2025comprehensive}. Their approach aimed to detect threats like the Follina vulnerability~\cite{regi2022case} by integrating multiple analytical techniques. The study showed that while AI and NLP are effective, they face challenges with unstructured data and high computational demands.

Malatji et al. examined the multifaceted dimensions of AI-driven cyberattacks, offering insights into their implications and mitigation strategies ~\cite{malatji2024artificial}. The study discussed the benefits and pitfalls of AI technologies in cybersecurity, emphasizing the need for robust defense mechanisms. A key limitation was the fast evolution of AI, outpacing current security measures and requiring constant adaptation.

Roshanaei et al. explored strategies for enhancing cybersecurity through AI and machine learning ~\cite{roshanaei2024enhancing}. The study emphasized real-time threat detection and predictive capabilities, highlighting the importance of integrating AI into security operations. However, the authors emphasized the ethical and legal concerns associated with AI deployment, highlighting the potential risks of over-reliance on automated systems without adequate human oversight.

Schmitt focused on AI-based cyber threat detection to protect digital infrastructures~\cite{schmitt2023securing}. The research evaluated machine learning classifiers for anomaly-based malware and intrusion detection, addressing challenges in deploying AI-enabled cybersecurity solutions. While the study provided valuable insights into integrating AI into security systems, it also recognized the limitations in adapting these models to the dynamic nature of cyber threats and the need for continuous learning mechanisms.

Based on the literature review presented above, it is evident that most existing research focuses on utilizing machine learning models to detect and analyze threats across various systems. To the best of our knowledge, there has been no prior work specifically dedicated to developing an AI agent for automated threat modeling. Therefore, our study is the first to introduce an agentic AI approach tailored for threat modeling, to enhance cybersecurity and ensure public safety.

%% file: methodology.tex
\section{Methodology}
\label{sec:methodology}
\begin{figure}[!ht]
    \includegraphics[width=1\columnwidth]{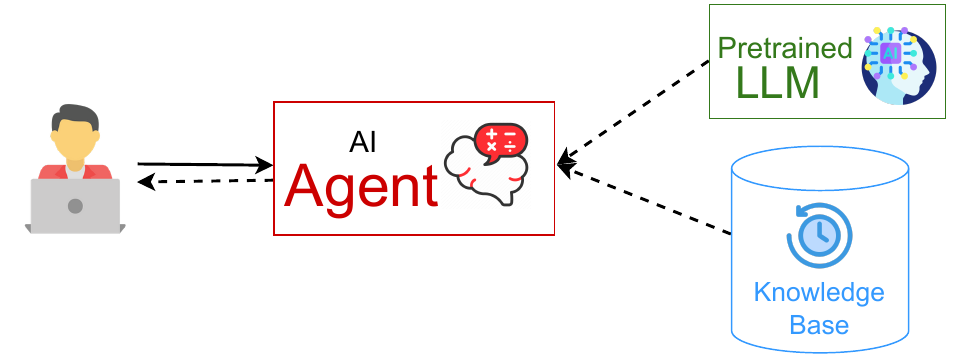}
    \vspace{-12pt}
    \centering
    \caption{ThreatGPT System Workflow}
    \label{fig_system_workflow}
    \vspace{-5pt}
\end{figure}

In this work, we designed and developed an Agentic AI system to facilitate intelligent, context-aware, and security-oriented query handling. Our proposed Agentic AI system comprises four primary layers: the Command-Line Interface (CLI) Layer, the AI Agent Layer, the Knowledge Base/Training Dataset Layer, and the Pretrained Large Language Model (LLM) Layer. The system's workflow is presented in Figure \ref{fig_system_workflow}.

\vspace{-12pt}

The CLI Layer provides a lightweight interface for users to efficiently submit queries and receive outputs. 

At the core of the system is the AI Agent Layer, which is responsible for managing the overall workflow. This layer handles the preprocessing of user inputs, determines the appropriate action strategy, formulates structured prompts for the language model, and postprocesses the outputs to ensure coherence, accuracy, and relevance. The agent dynamically decides whether to interact directly with the pretrained language model or to retrieve supplementary information from the knowledge base, depending on the nature of the query and the contextual requirements.

The Knowledge Base and Training Dataset Layer functions as a centralized repository of systematically compiled information, encompassing static datasets, user-specific data, and specialized resources such as threat modeling frameworks, including STRIDE, MITRE ATT\&CK, CVEs, NVD, and NIST standards. This layer plays a critical role in enhancing the system’s ability to ground the outputs of the large language model (LLM) in authoritative and domain-specific security knowledge, thereby improving the contextual relevance, accuracy, and reliability of the generated threat models.

\begin{figure}[!ht]
    \includegraphics[width=1\columnwidth]{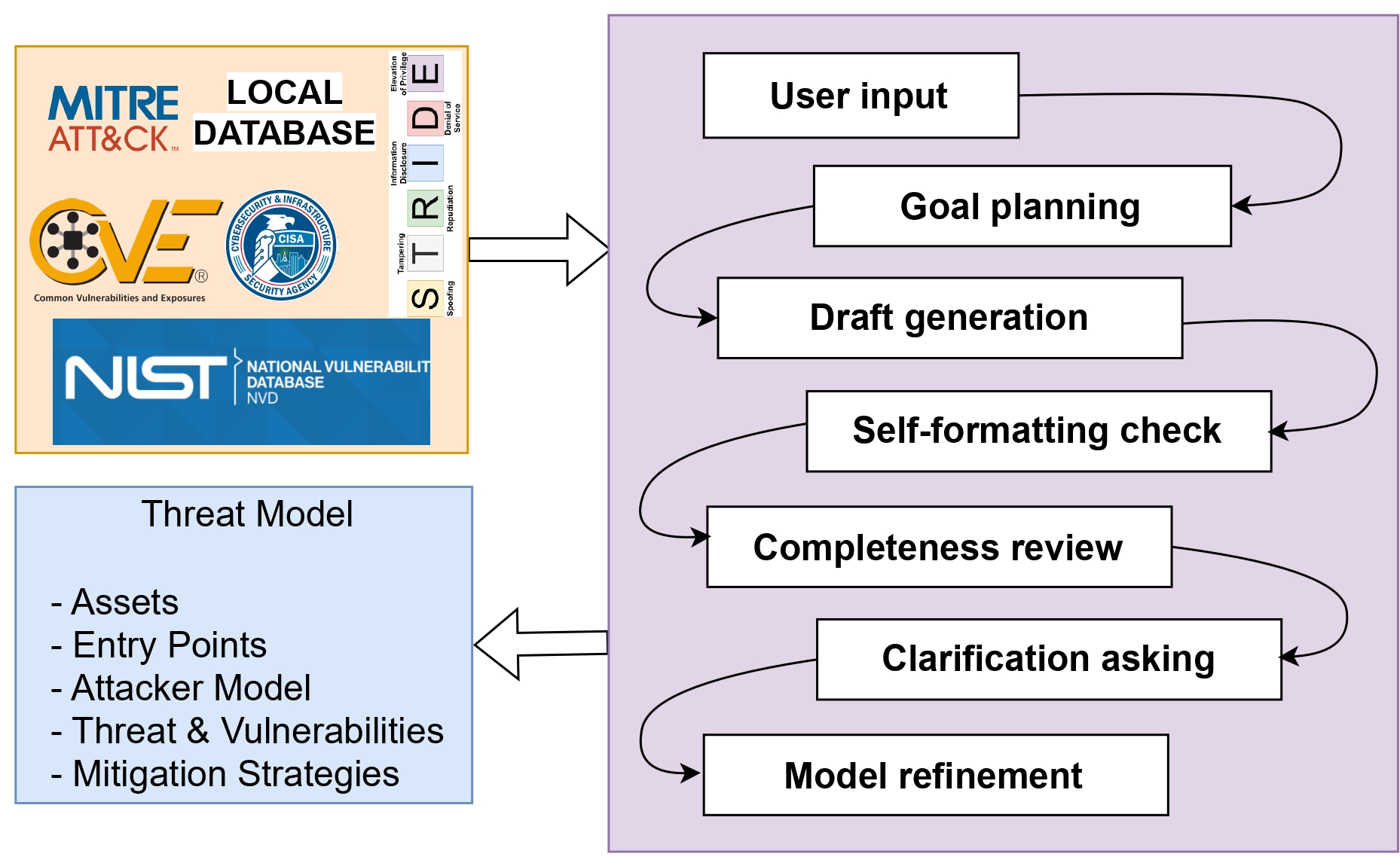}
    \vspace{-18pt}
    \centering
    \caption{System Architecture}
    \label{fig_system_architecture}
    \vspace{-8pt}
\end{figure}

The Pretrained LLM Layer integrates the Gemini API~\cite{team2023gemini}, a high-performance language model that generates natural language responses based on the agent's prompts. The system ensures high-quality, context-sensitive, and factually grounded output by using the Gemini model. The architecture of the system is presented in Figure \ref{fig_system_architecture}.

\vspace{-12pt}

To train the model effectively, we utilized a few-shot learning approach~\cite{parnami2022learning} by providing it with more than 50 detailed examples of threat models. These examples incorporate widely recognized cybersecurity frameworks and standards. Additionally, guidelines from respected sources like NIST and CISA were also integrated. This comprehensive training enabled the Gemini language model to better understand cybersecurity terms, threat categories, and real-world scenarios, resulting in improved accuracy and more context-specific, reliable responses.

When a user submits a request to generate a threat model for a specific system,  ThreatGPT initiates a systematic and autonomous workflow to generate high-quality threat models. It first recognizes its goal and develops a detailed multistep plan, encompassing asset identification, entry point analysis, threat mapping, vulnerability assessment, and mitigation suggestions. It proceeds to generate an initial threat model draft, ensuring adherence to the learned format by conducting an immediate self-verification and auto formatting phase if necessary. Following this, ThreatGPT performs an internal completeness review to identify any missing or ambiguous security aspects. If gaps are detected, it autonomously formulates clarification questions to obtain critical information from the user without interrupting the overall workflow. Once clarifications are received, ThreatGPT refines the initial model by integrating the new information, enhancing threat coverage across the frameworks it learned. Finally, it delivers the polished threat model, ensuring that the output is accurate, comprehensive, and professionally structured without requiring extensive human intervention.

%% file: experiments.tex
\section{Experiment and Results}
\label{sec:experiment}
The experiments were conducted using the free version of the Google Gemini API \cite{team2023gemini}, accessed through a local client setup. The system operated on an Ubuntu 20.04 LTS environment and was equipped with a Ryzen 7 5700U CPU with integrated graphics. The machine had 16 GB of RAM and a 1 TB SSD for storage. No external GPU acceleration was used. All client-side preprocessing, prompt construction, and postprocessing operations were performed using the local computational resources. The Gemini model itself remained hosted remotely on Google's servers, with local hardware primarily responsible for handling API interactions and lightweight processing tasks.

\begin{figure}[!ht]
    \includegraphics[width=1\columnwidth]{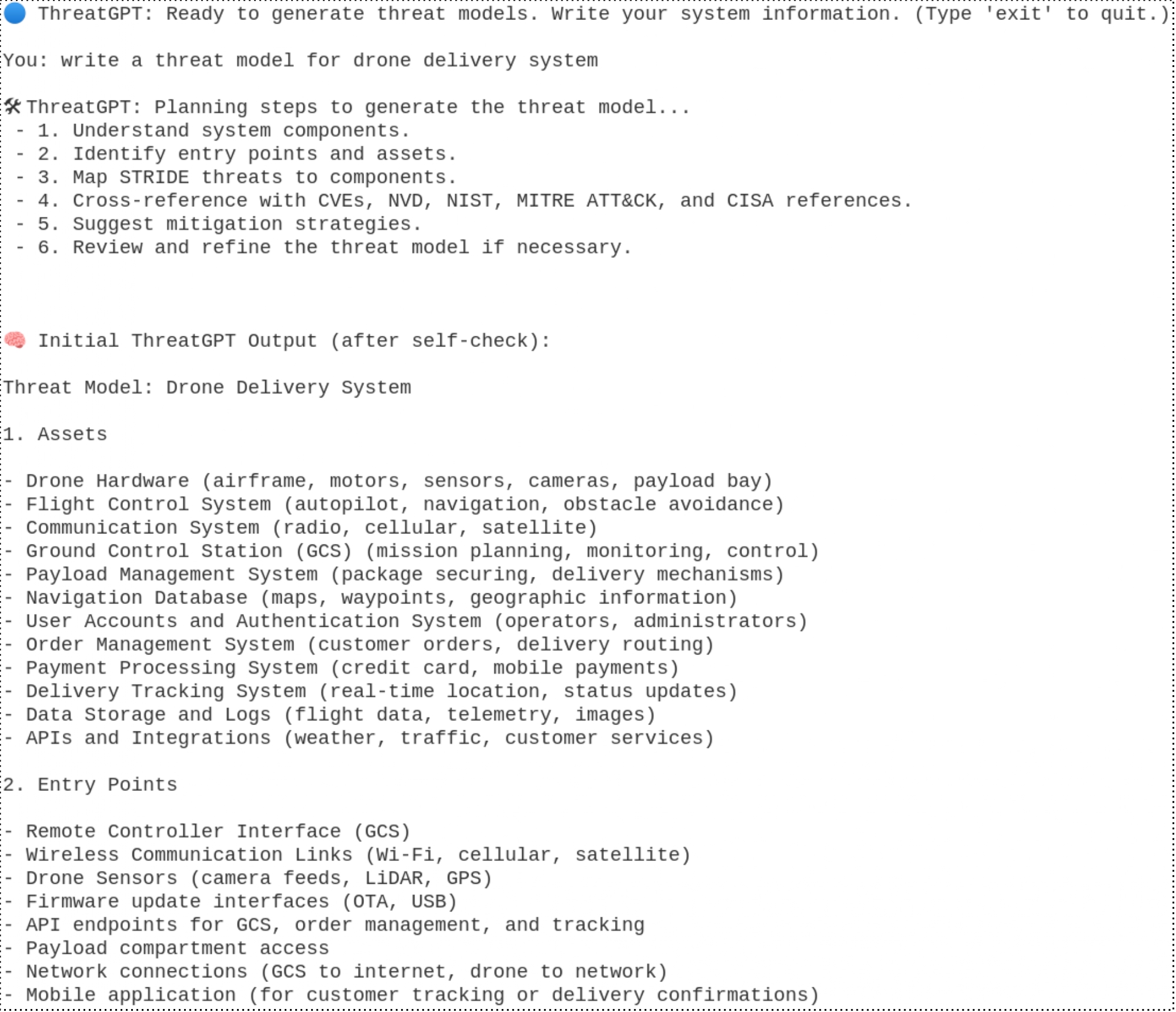}
    \centering
    \vspace{-15pt}
    \caption{Initial Threat Model}
    \label{fig_initial_threat_model}
\vspace{-8pt}
\end{figure}

\begin{figure}[!ht]
    \includegraphics[width=1\columnwidth]{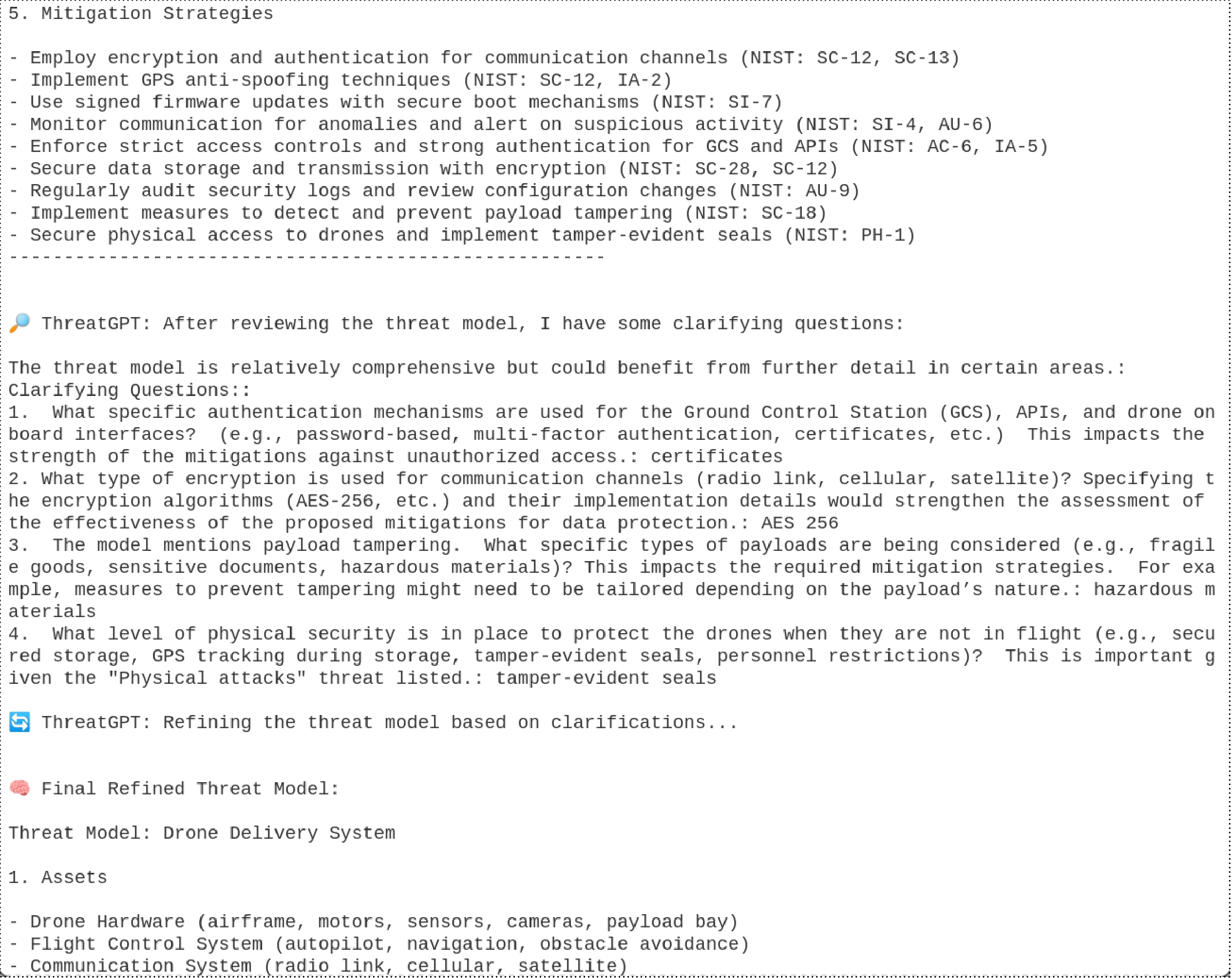}
    \vspace{-15pt}
    \centering
    \caption{Clarifying questions for refined threat model}
    \label{fig_refined_threat_model}
    \vspace{-20pt}
\end{figure}

The prompt and corresponding response samples generated by the AI model are presented in Figures \ref{fig_initial_threat_model} and \ref{fig_refined_threat_model}. The AI model systematically identifies and organizes key components of the threat model, including Assets, Entry Points, Attacker Models, Threats and Vulnerabilities, and Mitigation Strategies for a given system. ThreatGPT first generates an initial threat model. Then it asks for clarification questions if it needs to refine the model. For illustration purposes, only the important portions of the complete response are shown in the figures.

\vspace{-8pt}

We evaluated our model using three prompts. The simple prompt consists of the system name along with the detailed component names. It makes it easier for the model to understand and create the threat model. It does not need to think about the components. The compound prompt consists of the system name and some component names. The model needs to think about the detailed components and make a threat model for them. Finally, the complex prompt has only the system name. The model thinks about all the components and their interconnection to make the threat model. The sample prompts we used in each category are as follows:
\smallskip

\textbf{Simple Prompt:}
\textit{Generate a threat model for a Drone Delivery Management System where customer orders are processed through a mobile application connected to a scheduling server. The server assigns delivery tasks to drones, which navigate autonomously using a GPS-based navigation system. Package status is updated in real time via the tracking module, and operational data from drones is continuously monitored through a cloud-based dashboard. Consider the interactions and data flow between these components when identifying assets, entry points, threats, vulnerabilities, attacker models, and mitigation.}

\textbf{Compound Prompt:}
\textit{Generate a threat model for a Drone Delivery Management System. The main components include the drone navigation system, delivery scheduling server, package tracking module, customer mobile application, and cloud-based monitoring dashboard.}

\textbf{Complex Prompt:}
\textit{Generate a threat model for a Drone Delivery Management System.}
\smallskip

\begin{figure}[!ht]
    \includegraphics[width=1\columnwidth]{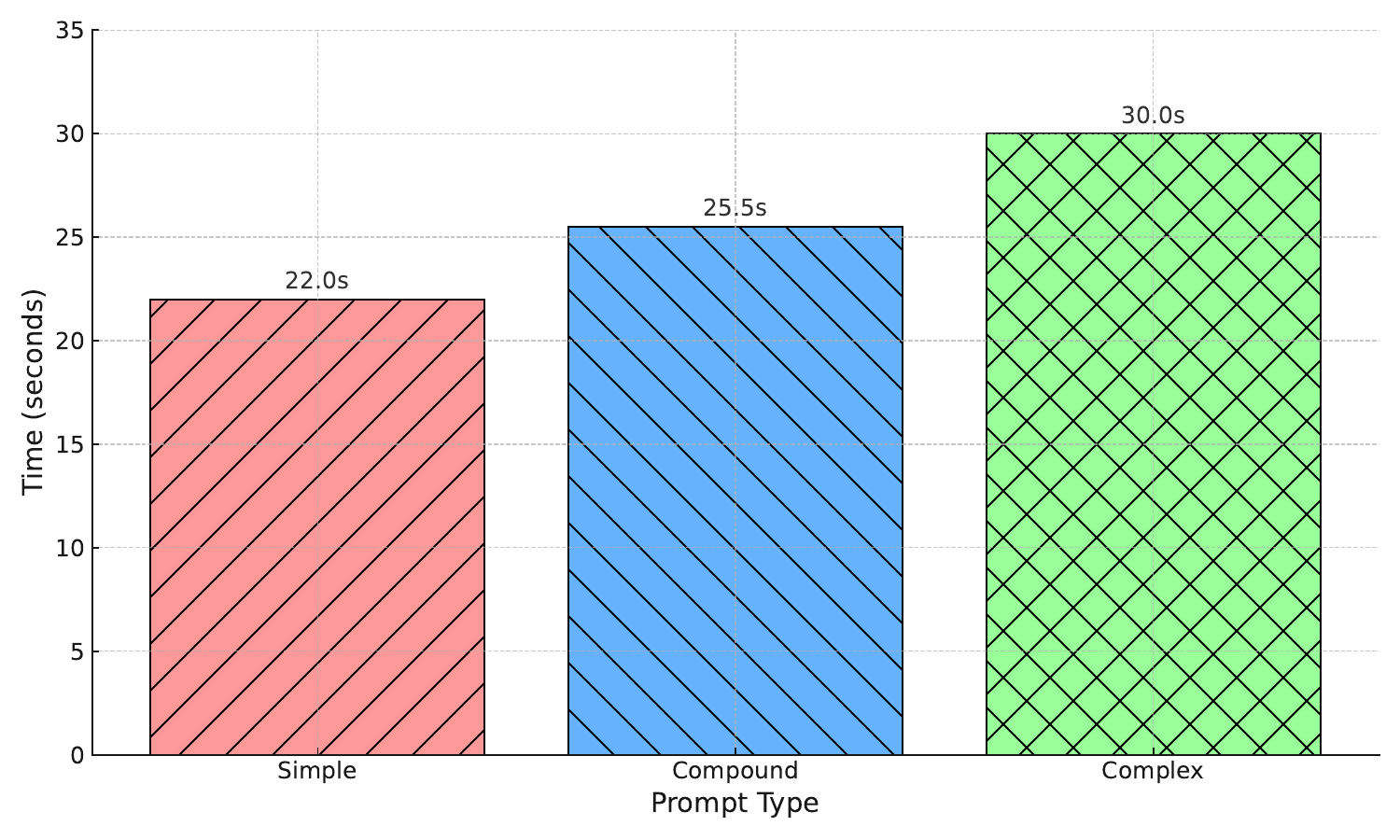}
    \centering    
    \vspace{-25pt}
    \caption{Response time comparison}
    \label{fig_response_type}
    \vspace{-15pt}
\end{figure}    

The time consumption for these three prompts is presented in Figure \ref{fig_response_type}.
Based on the figure, it is observed that the response time for each type of prompt remains within the range of 20 to 30 seconds. According to industry experts, the average time required for a human to generate a threat model for a specific system is over 40 hours \cite{Security_Compass_2025}, whereas our proposed model achieves this task within 30 seconds. Therefore, it can be concluded that our approach significantly reduces the time needed for threat model generation, offering a highly efficient alternative to manual methods. This rapid response capability is particularly critical for enhancing public safety, where timely identification and mitigation of threats are essential to protecting critical infrastructure and ensuring the security of communities.

%% file: conclusion.tex
\section{Conclusion}
\label{sec:conclusion}
In this study, we introduced ThreatGPT, an Agentic AI system designed to automate threat model generation using a few-shot learning approach with the Google Gemini API. By integrating over 50 compiled examples based on frameworks such as STRIDE, MITRE ATT\&CK, CVE, NIST, CISA and NVD, ThreatGPT demonstrated strong performance in producing structured, contextually relevant, and accurate threat models. Experimental results show that ThreatGPT significantly reduces modeling time, achieving a generation time of approximately 30 seconds compared to the 40+ hours typically required by human experts. This rapid generation capability has important implications for public safety, where timely identification of risks is very important to protecting infrastructure, maintaining operational continuity, and enhancing community resilience.

Despite its advantages, the system presents certain limitations. Its outputs depend heavily on the quality and scope of the few-shot examples, and it may occasionally miss highly specific or emerging threats not represented in the training data. Furthermore, reliance on external API interactions can introduce latency, and the use of the free Gemini API limits access to advanced fine-tuning capabilities. Future work will focus on expanding the diversity of threat modeling examples, integrating retrieval-augmented generation (RAG) for real-time access to live vulnerability databases, and enabling interactive refinement of threat models. These enhancements aim to further strengthen the robustness, adaptability, and applicability of ThreatGPT across cybersecurity and public safety domains.

\balance